\documentclass{article}

\usepackage[nonatbib,preprint]{neurips_2024}


\usepackage[utf8]{inputenc} 
\usepackage[T1]{fontenc}    
\usepackage{hyperref}       
\usepackage{url}            
\usepackage{booktabs}       
\usepackage{amsfonts}       
\usepackage{nicefrac}       
\usepackage{microtype}      
\usepackage{xcolor}         

\title{Reverse Browser: Vector-Image-to-Code Generator}

\author{%
  \href{https://orcid.org/0009-0004-2428-405X}{\includegraphics[scale=0.06]{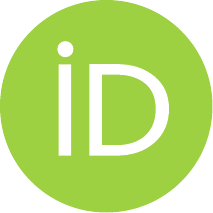}\hspace{1mm}Zoltan Toth-Czifra} \\
  University of Oxford \\
  \texttt{zoltan.toth-czifra@cs.ox.ac.uk} \\
  \texttt{tcz@tcz.hu} \\
}

\usepackage[style=numeric-comp, sorting=none, backend=biber, doi=true, isbn=false]{biblatex}

\usepackage{xcolor}

\usepackage{listings}
\usepackage{tikz}
\usetikzlibrary{shapes.misc, positioning}
\usepackage{collcell}
\usepackage{amsmath}
\usepackage{placeins}
\usepackage{pifont}
\usepackage{spverbatim}

\begin{document}

\maketitle

\begin{abstract}
  Automating the conversion of user interface design into code (image-to-code or image-to-UI) is an active area of software engineering research. However, the state-of-the-art solutions do not achieve high fidelity to the original design, as evidenced by benchmarks \cite{roberts_image2struct_2024, gui_webcode2m_2025}.

In this work, I approach the problem differently: I use vector images instead of bitmaps as model input. I create several large datasets for training machine learning models. I evaluate the available array of Image Quality Assessment (IQA) algorithms and introduce a new, multi-scale metric. I then train a large open-weights model and discuss its limitations.
\end{abstract}

\section{Introduction}

The World Wide Web is the most popular software development platform. A 2024 developer survey \cite{overflow_2024_2024} indicated that the top two languages used by professional software developers were JavaScript (62.3\%) and HTML+CSS (52.9\%). While building web interfaces is a visual task by nature, software developers spend considerable time implementing visual features through code, specifically HTML and CSS, as shown by the survey. 

Recent advances in deep learning, particularly the transformer architecture \cite{vaswani_attention_2023} have enabled rapid progress in powerful Artificial Intelligence (AI) models, in particular Foundation Models (FM), large-scale, versatile models trained on vast amounts of data. One of these models' most successful applications is coding assistants \cite{weisz_examining_2025}, which act as a conduit between the software engineer's intent and the resulting code. The primary modality of these assistants is text, while interface design is inherently a visual art, not text-based. \footnote{The Starry Night would surely be a less renowned painting if Vincent Van Gogh had made it by instructing an apprentice holding the paintbrush.} Automating code generation from UI design may enable software engineers to prototype quickly and designers to implement their creations without coding knowledge.

Several machine learning models have been proposed to solve image-to-code problems. The approach presented here differs from prior work in that it utilizes vector images as model input instead of bitmaps. Vector images may be more suitable for image-to-code tasks because they generally have lower intrinsic dimensionality than bitmap images of the same design and contain explicit structural information. The most popular design software used by UI designers can export designs as vector graphics, preserving explicit structural information.

\subsection{Contributions}

I implemented methods to produce large-scale training datasets from synthetic data and the public web. I developed and released a new IQA metric called Multi-Scale Pixel Similarity (MSPS). I performed proof-of-concept training experiments on synthetic data and smaller models, then fine-tuned and released the weights of a large vector-image-to-code model and analyzed its performance.

All the released software, datasets, and models are available under the Apache 2.0 license.

\section{Background}

In this context, (World Wide) Web refers to the application platform. It is a collection of technologies, principally the Hypertext Transfer Protocol (HTTP), the Hypertext Markup Language (HTML), Cascading Style Sheets (CSS), and JavaScript. In this paper, I focus solely on HTML\cite{web_hypertext_application_technology_working_group_html_2025} and CSS\cite{bos_cascading_2016}.

Computer graphics generally fit into two categories: bitmaps and vector graphics. A bitmap is a two-dimensional grid of pixel color values. A vector image, in contrast, is constructed using mathematical descriptions of shapes (paths, polygons, curves) as well as embedded text and images.

A transformer is an artificial neural network architecture introduced by \textcite{vaswani_attention_2023}. The self-attention mechanism in transformers can capture long-range dependencies between input and output sequences. Transformer-based models are often trained on large amounts of text (natural language and sometimes computer code) to perform language modeling tasks, predominantly text generation.

\subsection{Related work}
\label{sec: Related work}

Turning images into UI code using machine learning techniques is an active area of research, although no prior public works have used vector images as input.

\textbf{Pix2code} \cite{beltramelli_pix2code_2017} uses LSTM with convolutions to turn an input bitmap into HTML+CSS markup. It was trained on synthetic training data of relatively small cardinality. According to the author, the results are of low quality due to the model's small parameter space. The simple DSL\footnote{Domain-Specific Language} further limits the practical application. \textbf{Frontend Diffusion} \cite{ding_frontend_2025} works similarly: it builds a DSL document from hand-drawn sketches, which is used as input for a commercial Foundation Model to generate frontend code. The tool is not meant for precise design implementation. \textbf{Sketch2code} \cite{robinson_sketch2code_2019} turns bitmaps of hand-drawn wireframes into markup.  Using the approximate position of these elements, a simple HTML+CSS page can be generated. The resulting model is limited in terms of the design and the complexity of the output.
\textbf{Pix2Struct} \cite{lee_pix2struct_2023} is related but sets a different goal. It is a Visual Document Understanding model built to recognize the structure of documents such as receipts and business cards. It is an improvement over the previous state-of-the-art model, Donut \cite{kim_ocr-free_2022}. \textbf{ViCT} \cite{soselia_learning_2023} aims to turn web page screenshots or designs into HTML+CSS code. Similarly to Pix2Struct, it uses a multimodal transformer where the encoder leg is either a Vision Transformer (ViT) or DiT (Document Image Transformer), and the decoder leg is a text-based GPT-2 \cite{radford_language_2019} or Llama \cite{touvron_llama_2023} model. The authors note the model's limitations and admit that their training data "may not fully encapsulate the complexity of real-world web pages." A recent paper introduces \textbf{UICopilot} \cite{gui_uicopilot_2025}. The authors recognize the complexity of real-world web designs that previous works failed to address. They take design bitmaps as input and first extract the high-level document structure using a custom-trained vision transformer. They then use the extracted structure and the original bitmap as input for a commercial FM (GPT-4 \cite{openai_gpt-4_2024}) to produce the corresponding UI code.

A benchmark called \textbf{Image2Struct} \cite{roberts_image2struct_2024} aims to track the progress of the image-to-code capabilities of vision-language models. The authors remark that "we find that VLMs\footnote{Vision-language models, Foundation Models that were trained on both text and images.} are unable to perform well on our tasks".

\section{Data}
\label{sec: Data}

\subsection{Vector images}

The degrees of freedom in a bitmap file can be described by the formula \ref{eq:d_bitmap}.

\begin{gather} \label{eq:d_bitmap}
D_\text{bitmap} = W \times H \times C
\end{gather}

Here \(W\) and \(H\) are the width and height of the bitmap and pixels, and \(C\) is the number of color channels. The latter is typically 3, while the width and height may range from a few hundred to a few thousand pixels for a typical web design. In contrast, the degrees of freedom of a vector image may be described by formula \ref{eq:d_vector}.

\begin{align} \label{eq:d_vector}
D_\text{vector} &= \sum_{i=1}^{N} d_i = N \,\bar{d} \\
d_i &= \#\{\text{numeric parameters of primitive } i\} \nonumber \\
N &= \#\{\text{primitives}\} \nonumber \\
\bar{d} &= \text{mean } d_i \nonumber
\end{align}

A primitive is a mathematically described geometrical shape, such as a curve or a rectangle. For a typical web design, \(N \ll 1000\) and \(\bar{d} \ll 100\).

Images that are well-represented by geometric shapes, such as typical user interface designs, may be more compactly encoded by vector graphics rather than bitmaps. Furthermore, vector images contain explicit structure, unlike bitmaps, where the structure must be inferred. HTML has similar characteristics: a web page's Document Object Model (DOM) is a series of nested primitives and their attributes, further pointing to the use of vector images for image-to-code tasks. I chose Scalable Vector Graphics (SVG) \cite{amelia_bellamy-royds_scalable_nodate} for the vector image format.

\subsection{Datasets}

\begin{figure}[ht]
    \centering
    \centerline{\includegraphics[width=1.2\textwidth]{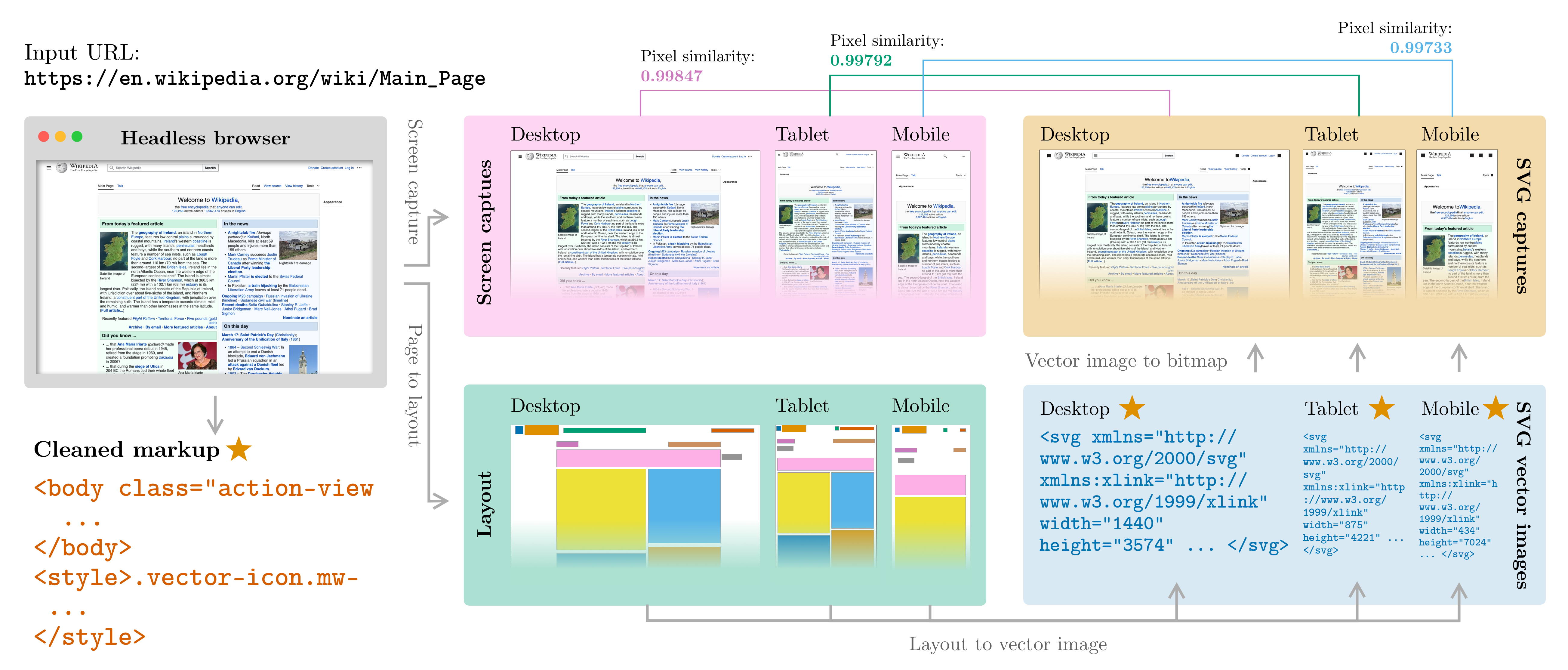}}
    \caption{Conceptual figure of rendering a web page into pairs of vector images and markup. Wikipedia's homepage is used as an example. The star symbol indicates the principal data included in the dataset. This example generates three different vector images for different viewport sizes.}
    \label{fig:rendering-flow}
\end{figure}

To obtain the SVG-markup\footnote{In this paper, I use "markup" to refer to both HTML and the corresponding CSS rules.} pairs to train a model on, the web page is first loaded and rendered in a headless browser engine\footnote{A web browser without a graphical user interface.}. Then, by iterating over the layout, a vector image can be constructed based on the attributes of each page element, such as position, size, color, text style, and so on. The resulting SVG file will represent the web page with a specific viewport size and will not dynamically reflow elements, unlike a web page. Figure \ref{fig:rendering-flow} explains the process visually.

Both the web pages and SVG files are then simplified and compressed without affecting the visual representation. Screen captures of both the original web page and the resulting vector image are saved for quality control.

The crawler was released under the Apache 2.0 license at \texttt{https://github.com/tcz/rb-crawler}

The crawler may be used to generate synthetic data from a set of artificially produced web pages. These may be generated from simple templates, probabilistic context-free grammar \cite{booth_applying_1973}, or even generative transformers. Synthetic data from simple web pages were used in several prior works mentioned in section \ref{sec: Related work}. As evidenced by these works, notably \textcite{soselia_learning_2023} and \textcite{beltramelli_pix2code_2017}, simple synthetic web pages will result in a low-complexity dataset and a \textit{low-perplexity} \cite{miaschi_what_2021} model with limited performance. In this work, several synthetic datasets were created for use in training proof-of-concept models. See Appendix \ref{app:synthetic-data}.

Training data obtained from the public web may result in a more versatile model due to the diversity and complexity of these web pages compared to a typical synthetic dataset, covering a wide range of web features.

URLs obtained from the Common Crawl \cite{common_crawl_common_2024} database were used to load and convert public websites to vector images. The items were then filtered by their similarity score, comparing the bitmap of the webpage with the bitmap rendering of the resulting vector image to filter out imperfect conversions. See section \ref{sec: Metrics} for more. The dataset contains $\approx 3.54\times10^{5}$ items and a total of $\approx 1.02\times10^{11}$ tokens, with a median token length of $\approx 1.51\times10^{5}$ per item.

Public web dataset was published under the Apache 2.0 License at \texttt{https://huggingface.co/datasets/tcz/rb-large}

\subsection{Data augmentation}

Data augmentation may multiply the number of training data instances obtained from a single web page. For example, randomly altering an already loaded web page's elements will result in a new item being created. While data availability has generally not been a concern when crawling the public web, data augmentation can provide a considerable speed-up, since augmenting an already loaded page is usually faster than crawling a new one.

Data augmentation may also be used to reduce the size of the markup and, consequently, the input sequence length used to train the model. This is important because in transformers, the time and space complexity of the self-attention mechanism scale quadratically \cite{tay_long_2020}.

The crawler implements several data augmenters, including two classes that aim to reduce the markup size using different strategies. New datasets produced with these size-reducing augmenters were made available at \texttt{https://huggingface.co/datasets/tcz/rb-large-reduced} and \texttt{https://huggingface.co/datasets/tcz/rb-large-chopped}

\section{Metrics}
\label{sec: Metrics}

Accuracy metrics measure visual differences between the input vector image and the output webpage. Designers and software engineers may expect no discernible difference between a design's bitmap representation and the corresponding web page's bitmap representation down to the level of individual pixels.

A reliable accuracy metric was required to assess the quality of markup to SVG conversion, to track the model's evolution during training, to evaluate the performance of a trained model, and to improve final model performance with best-of-N ranking \cite{snell_scaling_2024}.

Full-Reference Image Quality Assessment (FR-IQA) algorithms compare a reference bitmap image to a potentially degraded version. Not all FR-IQA algorithms are a good fit for the present use case. An ideal algorithm produces high loss with rotations, scaling, and large translations, but it is robust against small, few-pixel translations.

\subsection{Related work}

Works described in section \ref{sec: Related work} use various metrics to evaluate the performance of their respective models.

\textcite{soselia_learning_2023} employ a code-based metric and two image-based metrics. They coin \textit{htmlBLEU}, a syntax-aware version of the BLEU-score \cite{papineni_bleu_2001}. They find that \textit{htmlBLEU} correlates with the image similarity metric. This is workable for their very simple training data, but it is doubtful that such code-based metrics remain useful for complex training data. The paper notes that diverse HTML/CSS implementations make code-based metrics vulnerable to Type II errors.

As for image-based metrics, \textcite{soselia_learning_2023} use simple MSE. They also introduce a metric based on the intersection between non-empty pixels in the reference and the tested image.

\textcite{robinson_sketch2code_2019} uses MSE and SSIM \cite{zhou_wang_image_2004} to evaluate their model performance. \textcite{gui_uicopilot_2025} use SSIM and CLIP\cite{radford_learning_2021}. The latter uses the cosine similarity of latent vectors of a multimodal (text and image) model to compare images.

\medskip

A relatively recent paper by \textcite{roberts_image2struct_2024} discusses the problem of benchmarking vision-language models on image-to-structure tasks, including generating web pages from images. The authors introduce a new dataset called \textbf{Image2Struct} and develop two new metrics: cosine similarity between the inception vectors (CIS) and earth mover's similarity (EMS). CIS is similar to LPIPS discussed below, but it only uses a single layer.

EMS is a more innovative metric: it is based on Earth Mover Distance \cite{rubner_metric_1998} (EMD), which, in its original form, discards spatial pixel information and compares pixel value probability distributions in two images. EMS extends EMD with spatial information by subdividing the images into patches and incorporating their spatial distances into the EMD cost. The metric performs well when parts of the reference image are translated by small distances on the test image. Notably, the metric discards color information by converting images to grayscale first. The metric is highly sensitive to the chosen patch size. Additionally, their implementation is notably slower\footnote{Mean wall time to compare a pair of 1920×1080 pixels images: 5.31 seconds in EMS, 0.45 seconds in LPIPS, 0.13 seconds in MSPS} than most other FR-IQA metrics.

In addition to the two new metrics, the Image2Struct benchmark also calculates other metrics, including pixel similarity, SSIM, and LPIPS.

{
    \definecolor{gradred}{HTML}{EAAE7F}
    \definecolor{gradyellow}{HTML}{F5F099}
    \definecolor{gradgreen}{HTML}{7FCFB9}

    \newcommand*{\MinNumber}{0.0}%
    \newcommand*{\MidNumber}{0.5} %
    \newcommand*{\MaxNumber}{1.0}%

    \newcommand{\ApplyGradient}[1]{%
            \ifdim #1 pt > \MidNumber pt
                \pgfmathsetmacro{\PercentColor}{max(min(100.0*(#1 - \MidNumber)/(\MaxNumber-\MidNumber),100.0),0.00)} %
                \colorbox{gradred!\PercentColor!gradyellow}{\parbox{2cm}{\centering#1}}
            \else
                \pgfmathsetmacro{\PercentColor}{max(min(100.0*(\MidNumber - #1)/(\MidNumber-\MinNumber),100.0),0.00)} %
                \colorbox{gradgreen!\PercentColor!gradyellow}{\parbox{2cm}{\centering#1}}
            \fi
    }

    \newcolumntype{R}{>{\collectcell\ApplyGradient}c<{\endcollectcell}}
    \renewcommand{\arraystretch}{0}

    \begin{table}[ht]
    
    \centering
    \centerline{\resizebox{1.2\textwidth}{!}{
        \begin{tabular}{lRRRRRr}

            \fbox{\includegraphics[width=2cm]{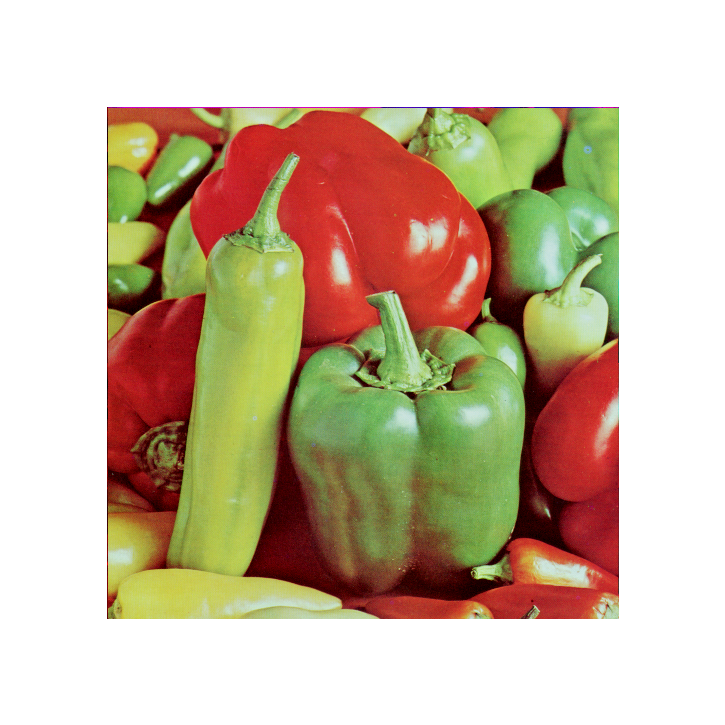}} &

            \multicolumn{1}{c}{\fbox{\includegraphics[width=2cm]{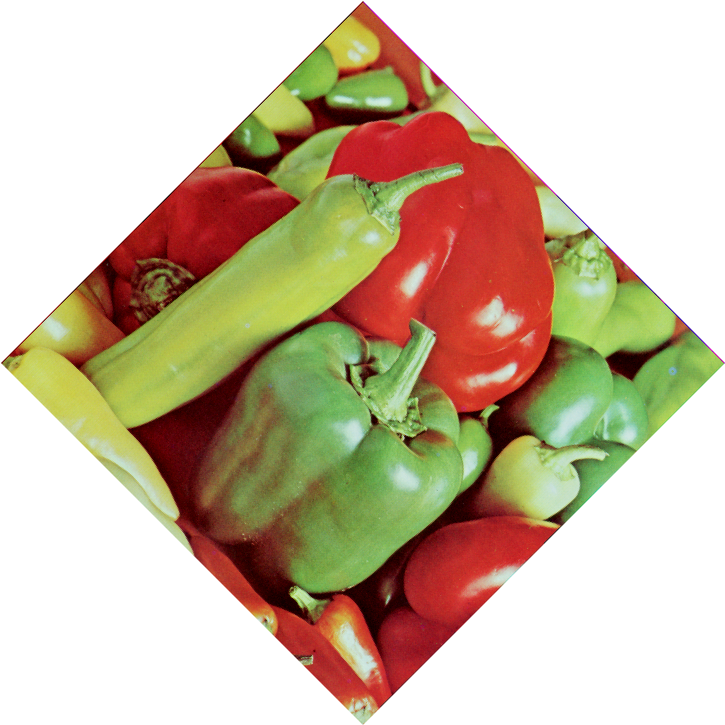}}} &

            \multicolumn{1}{c}{\fbox{\includegraphics[width=2cm]{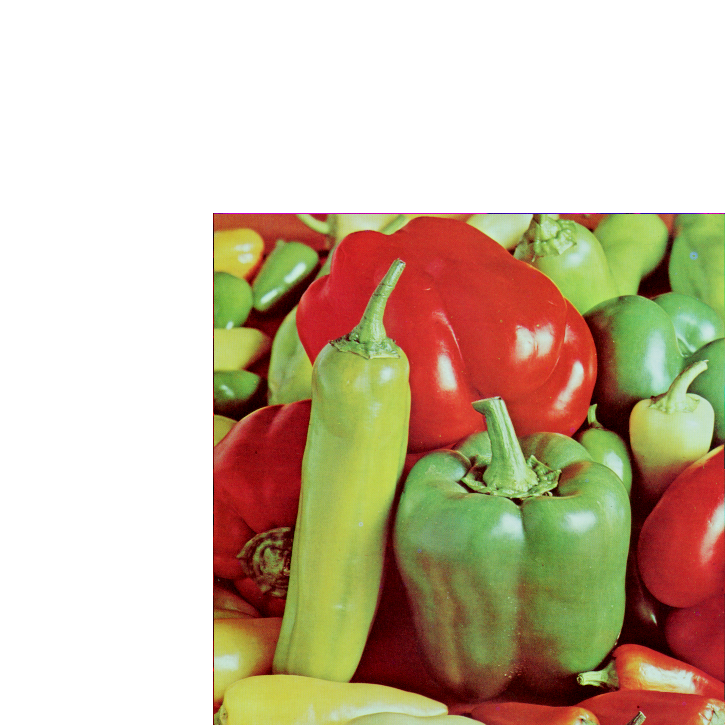}}} &

            \multicolumn{1}{c}{\fbox{\includegraphics[width=2cm]{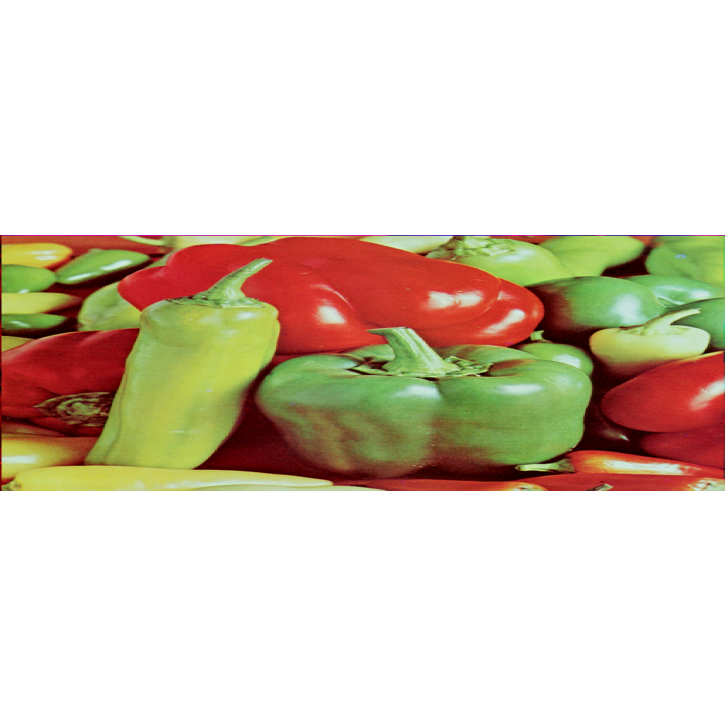}}} &

            \multicolumn{1}{c}{\fbox{\includegraphics[width=2cm]{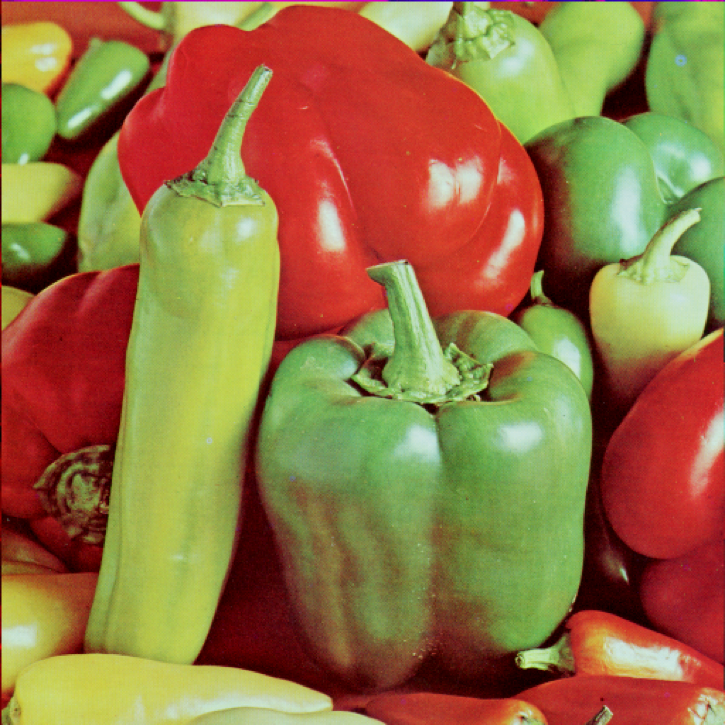}}} &

            \multicolumn{1}{c}{\fbox{\includegraphics[width=2cm]{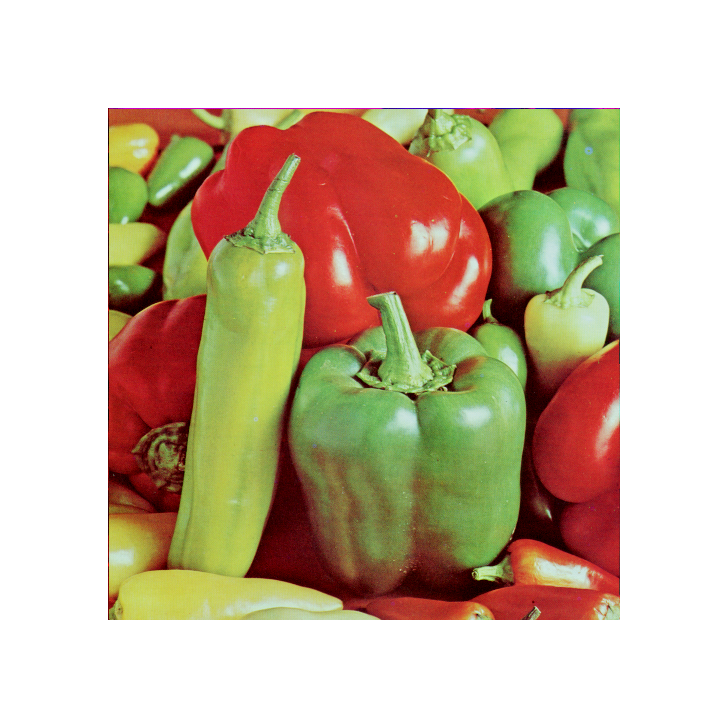}}} &

            \\

            \multicolumn{1}{c}{Reference} &
            \multicolumn{1}{c}{Rotated} &
            \multicolumn{1}{c}{Translated} &
            \multicolumn{1}{c}{Squeezed} &
            \multicolumn{1}{c}{Scaled} &
            \multicolumn{1}{c}{1px-trans.} &
            \multicolumn{1}{r}{Time (s)}

            \\ \hline

            TOPIQ \cite{chen_topiq_2023} & 0.7438108 & 0.7448106 & 0.7420042 & 0.7495018 & 0.1015147 & 0.247
            \\
            AHIQ \cite{lao_attentions_2022} & 0.9514484 & 1.0921408 & 1.0229274 & 0.9072852 & 0.5039899 & 0.269
            \\
            PieAPP \cite{prashnani_pieapp_2018} & 0.7264122 & 0.7321822 & 0.5625778 & 0.8485057 & 0.060036 & 0.146
            \\
            LPIPS \cite{zhang_unreasonable_2018} & 0.6214625 & 0.7138035 & 0.6818361 & 0.8698413 & 0.0407809 & 0.108
            \\
            DISTS \cite{ding_image_2020} & 0.1754472 & 0.1252416 & 0.254124 & 0.2622414 & 0.0069962 & 0.110
            \\
            WaDIQaM \cite{bosse_deep_2018} & 0.2324741 & 0.2426802 & 0.2575099 & 0.2473186 & 0.1206883 & 0.140
            \\
            CKDN \cite{zheng_learning_2021} & 0.6892447 & 0.6187835 & 0.6525854 & 0.6210397 & 0.3293804 & 0.121
            \\
            FSIM \cite{lin_zhang_fsim_2011} & 0.4262481 & 0.484176 & 0.4085104 & 0.5063123 & 0.0601805 & 0.248
            \\
            MS-SSIM \cite{wang_multi-scale_2003} & 0.6318116 & 0.7013864 & 0.6851234 & 0.7607858 & 0.0339208 & 0.249
            \\
            CW-SSIM \cite{sampat_complex_2009} & 0.7225356 & 0.7532272 & 0.7921229 & 0.6809691 & 0.0026404 & 1.567
            \\
            VIF \cite{sheikh_image_2006} & 0.9844843 & 0.9837387 & 0.9828228 & 0.9838052 & 0.7154267 & 0.282
            \\
            GMSD \cite{xue_gradient_2014} & 0.3110082 & 0.3265839 & 0.3245934 & 0.3145258 & 0.1130009 & 0.102
            \\
            NLPD \cite{laparra_perceptual_2016} & 0.9626138 & 0.9904734 & 0.9079056 & 1.091566 & 0.2708094 & 0.182
            \\
            VSI \cite{zhang_vsi_2014} & 0.2047782 & 0.2268612 & 0.2288077 & 0.2067673 & 0.0134498 & 0.201

        \end{tabular}
    }}
    \caption{Comparing several popular FR-IQA algorithms. A reference image was transformed in various ways, and the loss between the reference and the transformed image was calculated. \textit{1px-trans.} means that the reference image was translated by one pixel to the right and one pixel to the bottom. All loss scores were normalized to 0.0-1.0 where possible and standardized to mean "less is more similar." The reference image was taken from the \textcite{weber_usc-sipi_2018} dataset. The time of a single comparison was calculated on a machine with an Nvidia H100 GPU.}
    \label{tab:fr-iqa-algorithms}
    \end{table}

}

For this project, a range of popular FR-IQA algorithms was evaluated. Table \ref{tab:fr-iqa-algorithms} compares them in terms of their behavior across image transformations.

Learned Perceptual Image Patch Similarity (LPIPS) \cite{zhang_unreasonable_2018} was chosen as the fastest among those fitting the ideal criteria. LPIPS leverages deep features from convolutional neural networks (CNNs) trained for image classification. It calculates the distance between embedding vectors from the convolutional layers of the model. These distances are then passed through a model that converts them into human perceptual similarity scores.

\subsection{Multi-Scale Pixel Similarity}

\begin{figure}[t]
    \centering
    \centerline{\includegraphics[width=1.2\textwidth]{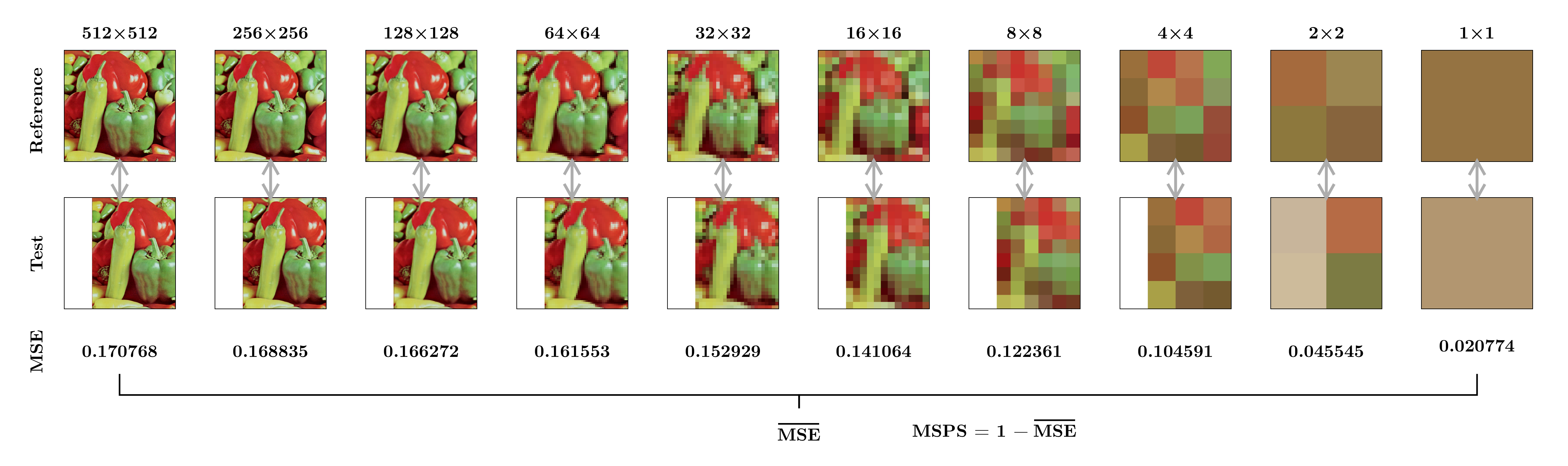}}
    \caption{Visualization of Multi-Scale Pixel Similarity (MSPS).}
    \label{fig:msps}
\end{figure}

LPIPS is a suitable choice for tracking training loss, evaluating model performance, and best-of-N ranking. However, a stricter metric was needed to assess the quality of the conversion from markup to SVG during dataset generation. FR-IQA algorithms are typically developed to evaluate image compression or restoration, considering the idiosyncrasies of the human visual perception system. They may overlook minor visual differences.

\medskip

To filter imperfect HTML-to-SVG conversion results, a new, stricter metric was developed, based on pixel-level comparison. It is called \textit{Multi-Scale Pixel Similarity} (MSPS).

MSPS uses the mean squared error (MSE) of pixel values and progressively reduces the sampling density of the input image, a similar approach to \textcite{wang_multi-scale_2003}. The final score is made up of the mean of all the MSEs at each density. See Figure \ref{fig:msps} for a visual explanation and equation \ref{eq:pixel-similarity} for the formula.

\begin{multline} \label{eq:pixel-similarity}
  \text{MSPS}\bigl(I_1, I_2\bigr)
  \;=\; 1
  \;-\; \\
  \frac{1}{N} \sum_{i=1}^{N}
  \left[
    \frac{1}{H_i W_i C}
    \sum_{h=1}^{H_i} \sum_{w=1}^{W_i} \sum_{c=1}^{C}
    \Bigl( V_1^{(i)}(h, w, c) - V_2^{(i)}(h, w, c) \Bigr)^2
  \right]
\end{multline}

Here, \(I_1\) and \(I_2\) are the input images to be compared. \(V_1^{(i)}\) and \(V_2^{(i)}\) refer to the pixel value at location \((h, w)\), channel \(c\) in a scaled down version of images \(I_1\) and \(I_2\) by factor \(i\). For instance, \(V_1^{(1)}\) and \(V_2^{(1)}\) refer to the pixel values of the images without scaling, while \(V_1^{(2)}\) and \(V_2^{(2)}\) are half the size of the originals, etc. \(H\), \(W\), and \(C\) respectively refer to the pixel height, pixel width, and the number of channels of the compared images at scale \(i\). Scaling is done on both axes using 2D average-pooling.

\(N\) is the total number of MSE calculations and can be calculated like so:

\begin{gather} \label{eq:pixel-similarity-n}
N \;=\; 1 \;+\; \left\lfloor \log_{2}\!\Bigl(\min(H_1, W_1)\Bigr) \right\rfloor
\end{gather}

Here, \(H_1\) and \(W_1\) refer to the pixel dimensions of the original image.

MSPS and LPIPS are, expectedly, moderately correlated (Pearson correlation of $\approx -0.62$ with p-value of \(1.19 \times 10^{-95}\) on the Image2struct dataset). However, MSPS is a lower-level metric that ignores the characteristics of human vision and ensures strict quality assessment of the web-to-vector conversion while remaining robust against small translations. I released an implementation of MSPS under the Apache 2.0 License at \texttt{https://github.com/tcz/rb-msps}

\section{Model training}

\subsection{Evaluation}

Token‑level cross‑entropy loss does not give useful insights into the model's performance. To monitor the domain-specific performance during training, after each \(N\) steps, the model was evaluated on the accuracy metrics discussed in Section \ref{sec: Metrics}, LPIPS, and MSPS. The generated markup was rendered in a headless browser, and screenshots of the predicted and expected markups were taken for comparison to produce accuracy scores.

This is a relatively slow process compared to traditional sequence-to-sequence evaluation metrics, such as BLEU, which limits the practical size of the evaluation data split.

\subsection{Proof-of-concept}

Calculating the layout of a web page involves a nested series of simple arithmetic operations, such as additions, multiplications, and divisions. The trained model is expected to perform the inverse of the layout operations and produce the corresponding HTML and CSS from an image resembling the layout. Transformer models can learn to perform arithmetic operations \cite{mcleish_transformers_2024, lee_teaching_2023, zhou_pre-trained_2024}, but with limited accuracy and complexity. Before training larger models, the concept was evaluated by pre-training or fine-tuning smaller models with simple but progressively more complex synthetic data. 

For these experiments, the T5 family (T5 and LongT5) \cite{raffel_exploring_2023, guo_longt5_2022} was selected, which is a set of pre-trained encoder-decoder models. The model performed well on synthetic data, see Appendix \ref{app:synthetic-data} for results. Figure \ref{fig:experiment-analysis-longt5-color-text-size} shows pre-training and fine-tuning results on a LongT5 Base model using synthetic data.

\begin{figure}[h]
    \centering
    
    \begin{subfigure}{\textwidth}
    \centerline{\includegraphics[width=1.2\textwidth]{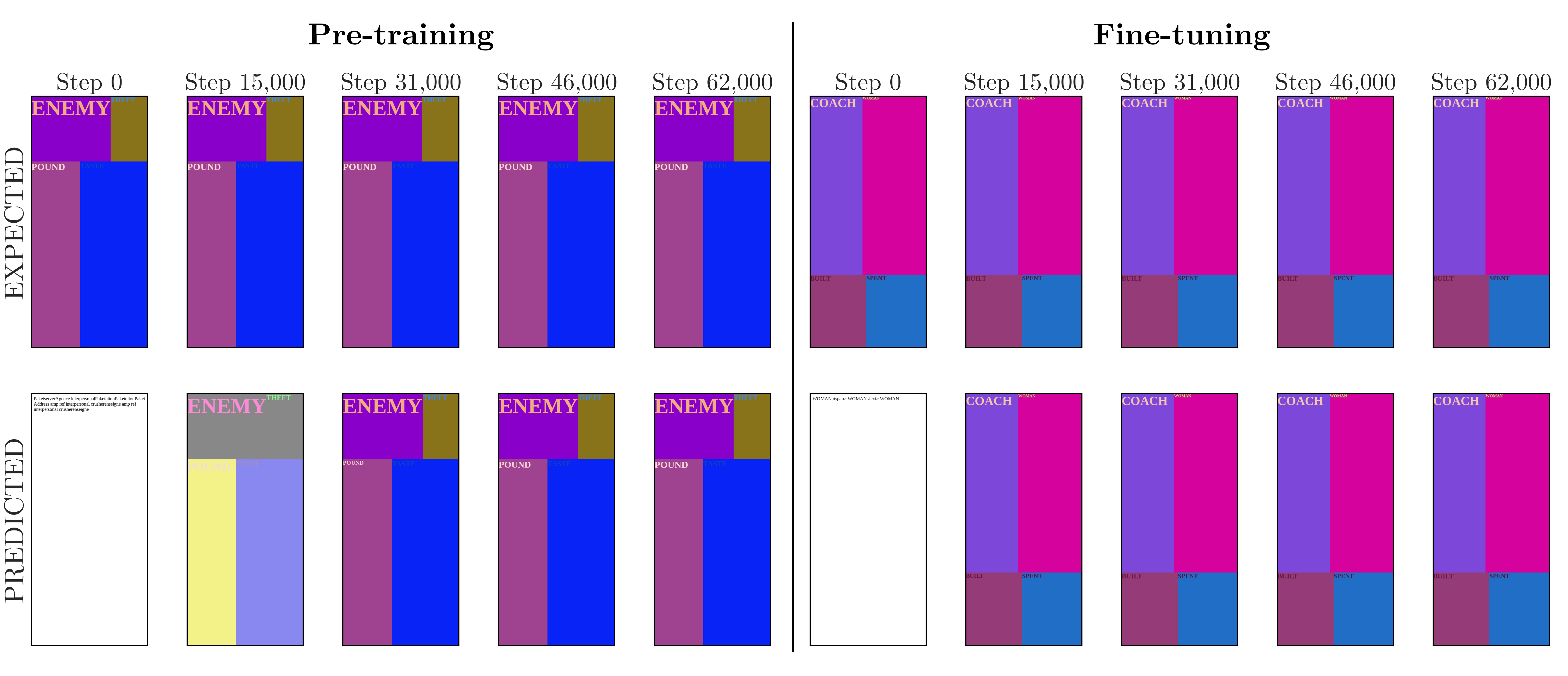}}
    \caption{Combined evolution of experiment predictions}
    \end{subfigure}
    
    \medskip
 
    \begin{subfigure}{\textwidth}
    \centerline{\includegraphics[width=1.2\textwidth]{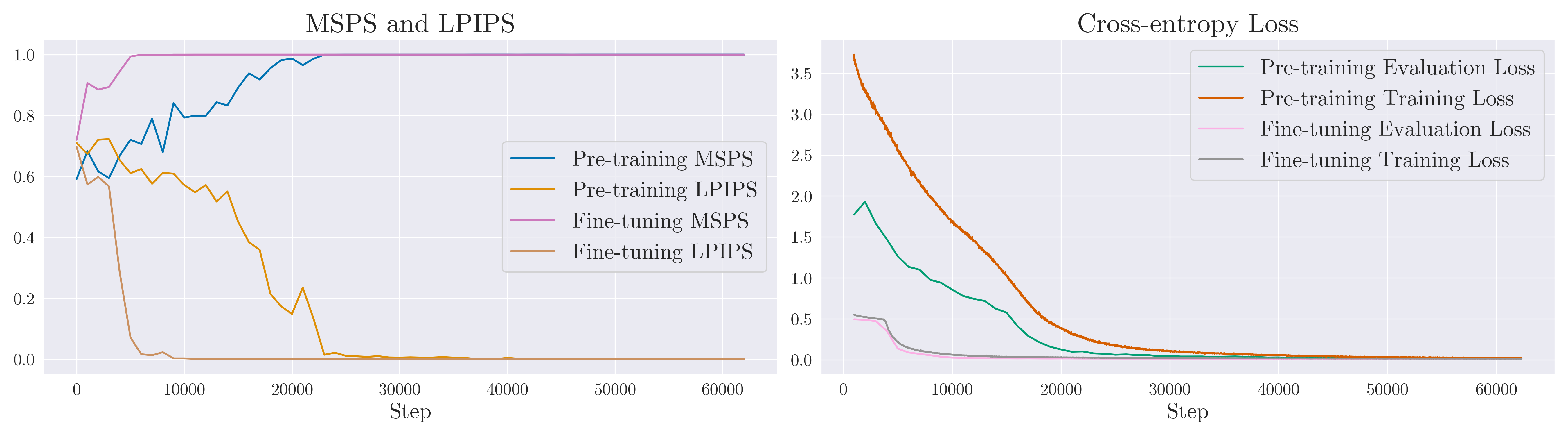}}
    \caption{Combined accuracy metrics and cross-entropy loss}
    \end{subfigure}
    
    \caption{LongT5 pre-training and fine-tuning on the "Color+Text+Size" dataset. The upper figure shows the evolution of experiment predictions using a single sample from the validation dataset, rendered in a browser engine, compared to the expected ground truth. The lower left figure shows the evolution of the accuracy metrics on the validation dataset as defined in Chapter \ref{sec: Metrics}. The lower right figure shows the cross-entropy loss on the validation (unseen) and the training datasets.}
    \label{fig:experiment-analysis-longt5-color-text-size}
\end{figure}
\FloatBarrier

\subsection{Larger models}

Training on complex, heterogeneous public web data exceeded the capacity of the T5 family both in terms of resulting accuracy and input length. The Llama 3.2 \cite{grattafiori_llama_2024} model family was chosen as a base to fine-tune on public data. It is a family of decoder-only models with a modern architecture, a wide range of model sizes,  broad framework availability, and a permissive license. It was pre-trained on 15 trillion multilingual tokens, including code and math through special pipelines. The fine-tuning was done using Low-Rank Adaptation (LoRA) \cite{hu_lora_2021} with quantized weights. For hyperparameters, see the published experiments.

After successfully fine-tuning the smallest (1B) model on simple synthetic data (0.9947 MSPS and 0.0100 LPIPS on the test split), a series of experiments was performed on increasingly larger variants of Llama (1B, 3B, 90B) using the public web dataset. Increasing the model size was necessary because the smaller models failed to produce satisfactory accuracy levels.

\begin{figure}[h]
    \centering
    \centerline{\includegraphics[width=1.2\textwidth]{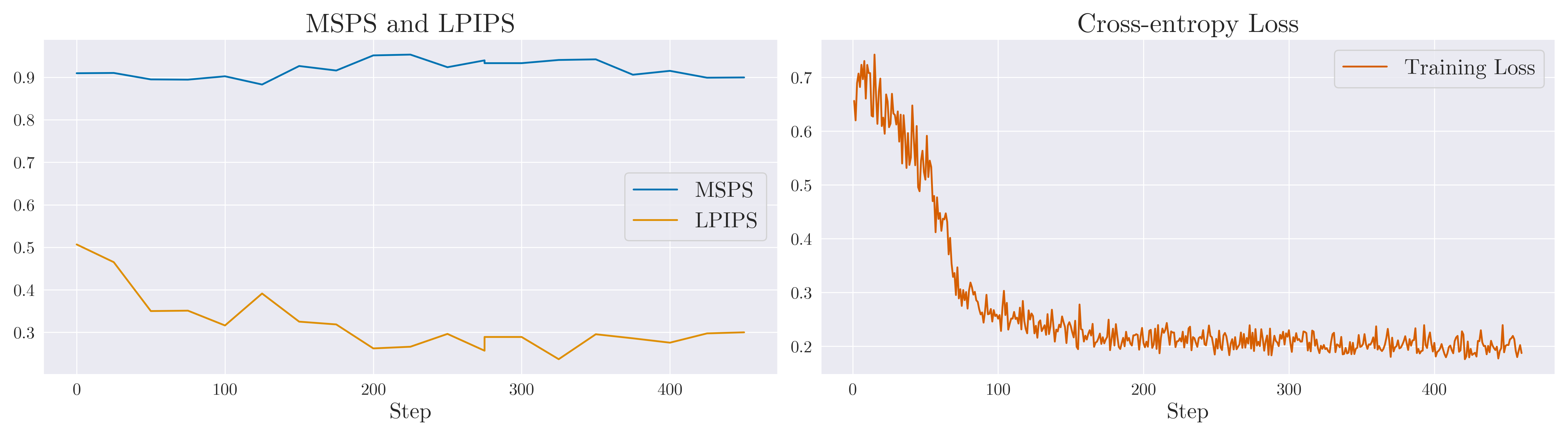}}
    \caption{Llama 3.2 90B fine-tuning on the "Large" dataset. The left-side chart shows the evolution of the accuracy metrics on the validation dataset as defined in \ref{sec: Metrics}. The right-side chart shows the cross-entropy loss on the training dataset. Due to an unsolved bug in the training framework, the validation cross-entropy was not logged.}
    \label{fig:experiment-analysis-llama90b-large}
\end{figure}

Training the largest, final model (90B) on a single epoch training using a length-filtered version of the "Large" public web dataset took over 392 hours (\(3.64 \times 10^{20}\) FLOPs) on a single H100 GPU. The training dataset had to be filtered to contain only items shorter than 12,000 tokens due to the memory limits of a single GPU. Figure \ref{fig:experiment-analysis-llama90b-large} shows the training loss evolution. The best model was saved at 325 steps (validation split MSPS: 0.9532, LPIPS: 0.2369).

\FloatBarrier

The final test set scores are: \textbf{0.9530 MSPS} and \textbf{0.3012} LPIPS. It is worth noting that the largest Llama 3.2 model produced markedly higher accuracy than the smaller variants, even before fine-tuning.

The documented experiments were released under the Apache 2.0 License at \texttt{https://github.com/tcz/rb-experiments}

\section{Results}

\subsection{Benchmark}
\label{sub: Benchmark}

No publicly available benchmark exists for models turning vector files into web pages.

However, there are two public benchmarks for bitmap-to-webpage tasks: Image2Struct \cite{roberts_image2struct_2024} and WebCode2M \cite{gui_webcode2m_2025}. Both test an array of multimodal FMs, including some of the most powerful commercial models, on datasets compiled by the authors. The tested FMs are not fine-tuned on domain-specific data. Only Image2struct publishes a regularly updated leaderboard\footnote{https://crfm.stanford.edu/helm/image2struct/latest/\#/leaderboard/image2webpage}.

\medskip

It must be stressed that the following is not an Image2Struct benchmark of the final model. Image2Struct assumes a distinct input (bitmaps), which is supposedly a more difficult task than vector-to-code and requires a more advanced visual model. Still, comparing accuracy scores of bitmap-to-code and vector-to-code tasks using the same input web pages can be insightful and underline that vector images might be a more appropriate choice of input for common software engineering scenarios.

Thus, I crawled Image2Struct's web page dataset with the tool introduced in Section \ref{sec: Data} to obtain SVG inputs for each web page. I then performed inference with them on the final model and calculated the LPIPS\footnote{Image2Struct authors normalize LPIPS to be a greater-is-better metric by subtracting the traditional LPIPS value from 1. They also use a different neural network to calculate LPIPS. They used VGG, whereas previously in this paper AlexNet was used. For the benchmarks, their parameters were followed.} and EMS (discussed in Section \ref{sec: Data}) scores for the generated HTML+CSS code. 

Image2Struct leaderboards prefer the latter metric, although they publish LPIPS scores as well. Table \ref{tab:image2struct-scores} shows the results.

\begin{table}[ht]
\centering
\centerline{\resizebox{1.2\textwidth}{!}{
    \begin{tabular}{lrrrrrr}
        \hline
        \textbf{Model} & EMS & \textbf{EMS (vector)} & LPIPS* & \textbf{LPIPS* (vector)} & \textbf{Compilation success} & \textbf{Training and inference success} \\ \hline
        GPT-4o (2024-08-06) & \textbf{0.735} &  & 0.018 &  & 0.998 &  \\
        Claude 3.5 Sonnet (20240620) & 0.724 &  & 0.285 &  & 0.972 & \\ 
        Claude 3.5 Sonnet (20241022) & 0.712 &  & 0.011 &  & 0.989 &  \\
        GPT-4o (2024-05-13) & 0.71 &  & 0.333 &  & 0.98 &  \\
        GPT-4o mini (2024-07-18) & 0.696 &  & 0.012 &  & 0.986 & \\ 
        Gemini 1.5 Pro (0409 preview) & 0.683 &  & 0.367 &  & 0.971 & \\ 
        Gemini 1.5 Pro (002) & 0.683 &  & 0.014 &  & 0.919 &  \\
        Gemini 1.5 Flash (002) & 0.674 &  & 0.008 &  & 0.916 &  \\
        GPT-4V (1106 preview) & 0.653 &  & 0.378 &  & 0.957 &  \\
        Claude 3 Sonnet (20240229) & 0.642 &  & \textbf{0.383} &  & 0.956 &  \\ 
        Mistral Pixtral (2409) & 0.642 &  & 0.015 &  & 0.921 &  \\
        Palmyra Vision 003 & 0.64 &  & 0.311 &  & 0.884 &  \\
        Gemini 1.0 Pro Vision & 0.502 &  & 0.316 &  & 0.795 &  \\
        LLaVA 1.6 (13B) & 0.447 &  & 0.332 &  & 0.731 &  \\
        LLaVA 1.5 (13B) & 0.435 &  & 0.379 &  & 0.773 &  \\
        Claude 3 Opus (20240229) & 0.378 &  & 0.319 &  & 0.655 &  \\
        Qwen-VL Chat & 0.008 &  & 0.017 &  & 0.031 &  \\
        IDEFICS-instruct (80B) & 0.001 &  & 0.001 &  & 0.002 & \\ 
        IDEFICS-instruct (9B) & 0.001 &  & 0.001 &  & 0.001 &  \\
        IDEFICS 2 (8B) & 0 &  & 0 &  & 0 &  \\ \hline
        Reverse Browser (proposed) &  & 0.693 &  & 0.755 &  & 0.981 \\ \hline
    \end{tabular}
}}
\caption{The performance of the final Reverse Browser model on a vector-to-webpage task, contrasted with multimodal FM performance on bitmap-to-webpage as defined by Image2Struct. Both tasks use the same set of web pages for ground truth, but the scores are not comparable because they measure the performance on different tasks. The two different success rates are related to different failure modes, hence the separation. The updated FM leaderboard was copied from the Image2Struct website on May 6, 2025.}
\caption*{\footnotesize * Image2Struct uses a normalized LPIPS where greater means better. They also use a neural network different from the one used in this paper (VGG over AlexNet). In this table, their metric is adopted.}
\label{tab:image2struct-scores}
\end{table}

\begin{figure}[h]
    \centering
    \centerline{\includegraphics[width=1.2\textwidth]{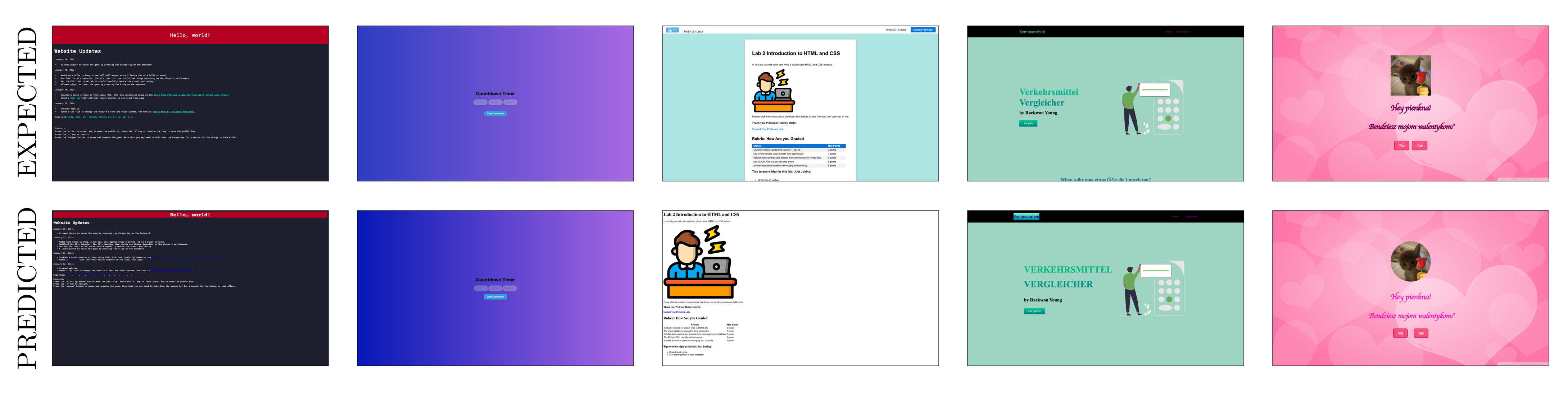}}
    \caption{A few randomly selected examples from the model's output on the Image2Struct web pages dataset.}
    \label{fig:image2struct-samples}
\end{figure}

Figure \ref{fig:image2struct-samples} shows a few examples of the model's output on the Image2Struct set of websites. In Appendix \ref{app:image2struct-more}, I write briefly on my methods of turning the web pages in the Image2Struct dataset into vector images and how inference was performed.

\FloatBarrier

\subsection{Use case}
\label{sub: Use case}

The test set accuracy scores of the final model (\textbf{0.9530} MSPS and \textbf{0.3012} LPIPS) suggest that it is not ready for use in a real-world software engineering workflow. Nevertheless, in this section, I conduct a brief qualitative test on the finished model and demonstrate a potential use case.

\begin{figure}[h]
    \centering
    \centerline{\includegraphics[width=1.2\textwidth]{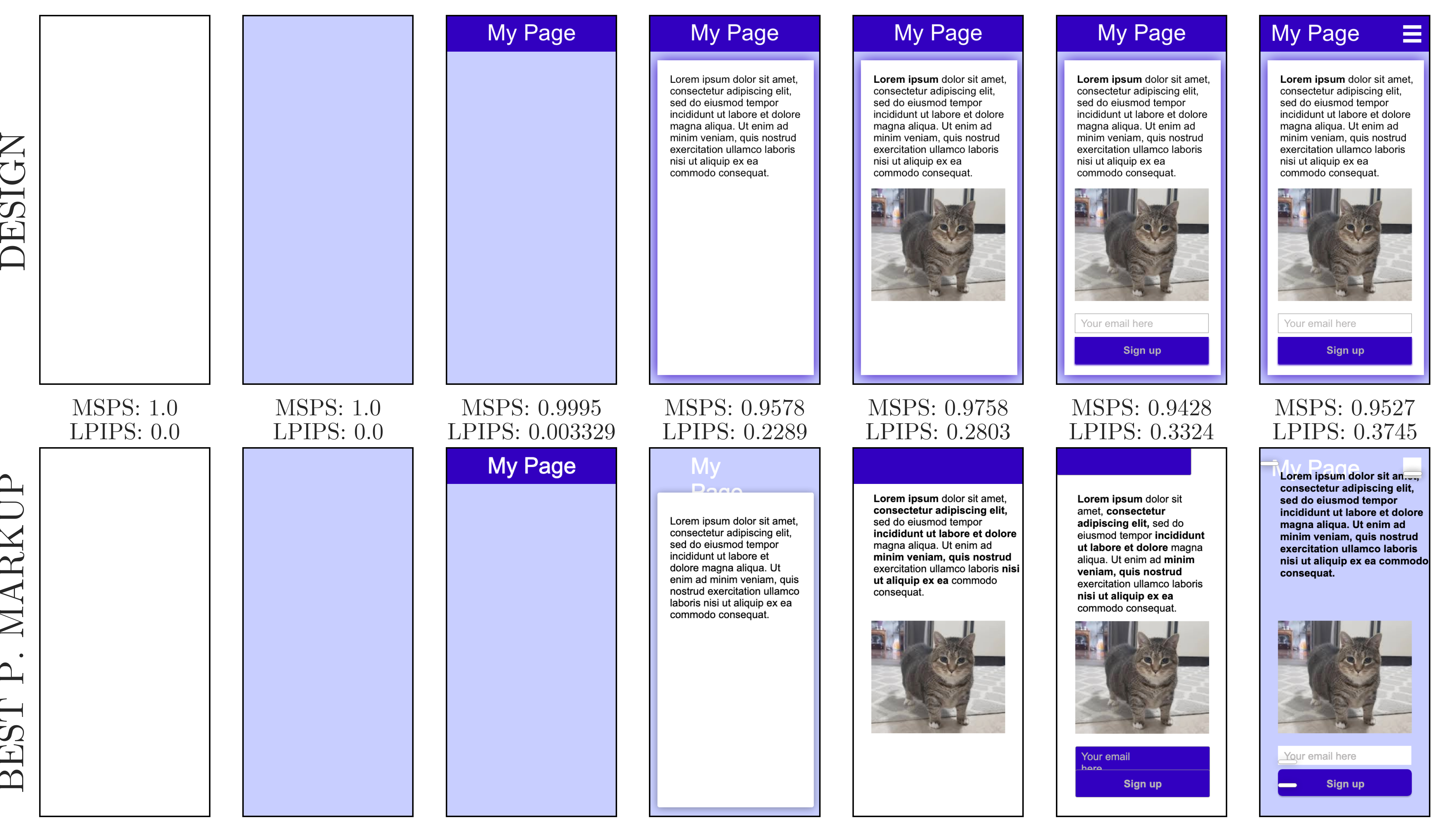}}
    \caption{The best model output by LPIPS out of 50 samples for each design step, with accuracy metrics between the original design and the resulting web page screen capture. Here, generating \texttt{svg} elements was forbidden, and a lower temperature of 0.5 was used.}
    \label{fig:design-test-50-lower-temp}
\end{figure}

A software engineer may want to develop a simple web application. They start by designing the visual elements. They may use a prototyping software such as Sketch\footnote{https://www.sketch.com/} to progressively create the application's design. The top row of Figure \ref{fig:design-test-50-lower-temp} shows the imaginary steps of creating the design. Each step adds an element or makes a minor change. At each step, the software engineer may generate the markup and open it in a browser, iterating on the design and testing on different devices. When the design is ready, the generated markup can serve as a starting point for the HTML and CSS implementation.

\FloatBarrier

The bottom row shows the markup inferred from the design using the final model using best-of-N sampling \cite{snell_scaling_2024} (samples: 50, temperature: 0.5, top-p: 0.9) along with the accuracy score. Notably, the SVG files exported from Sketch may have a different "flavor" from the ones we created for the training datasets.

\begin{figure}[h]
    \centering
    \centerline{\includegraphics[width=1.2\textwidth]{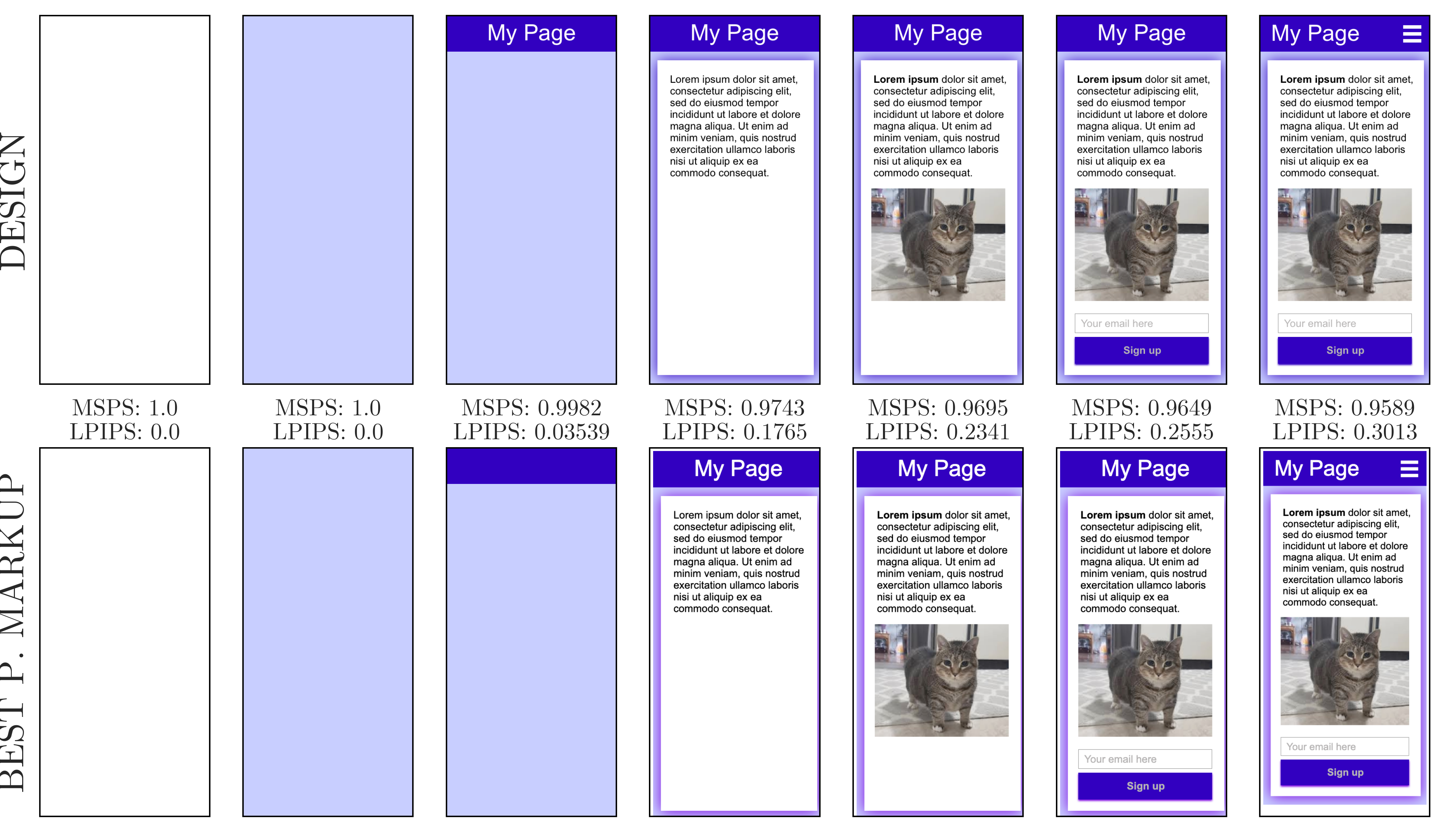}}
    \caption{The best model output by LPIPS out of 25 samples for each design step without filtering \texttt{<svg>} tags. The last pages show the results of reward hacking.}
    \label{fig:design-test-25-high-temp}
\end{figure}

The model is prone to reward hacking as demonstrated in Figure \ref{fig:design-test-25-high-temp}. In some cases, an SVG output was generated from the SVG input, which is a valid subset of HTML \cite[4.8.16 SVG]{web_hypertext_application_technology_working_group_html_2025}. For that reason, the generated markups were filtered for \texttt{<svg>} tags.

\FloatBarrier

\section{Discussion}

I interpret the Image2Struct results in \ref{sub: Benchmark} as follows:

\begin{enumerate}
    \item Even large, commercial, multimodal FMs perform poorly at image-to-UI tasks. The Image2Struct prompts contain additional input besides the bitmaps (extracted text, list of resources), yet the results are not usable in a real software engineering setting. More research and development are needed before the industry can adopt image-to-UI.
    \item A relatively small model trained by an individual with a limited budget can deliver accuracy scores similar (or, in the case of LPIPS, significantly better) to the best bitmap-to-webpage scores. This suggests that vector images may be a better choice for image-to-UI than bitmaps.
\end{enumerate}

\subsection{Limitations}

Despite promising experiments on synthetic data, the accuracy of the final model trained on complex public web data stays below what could be considered commercially viable.

Inference on the final model is relatively slow. The per-token speed ranges from $\approx1000$ to $\approx30$ tokens per second for ingestion and $\approx150$ to $\approx60$ tokens per second for generation on an Nvidia H100 GPU. Optimizing the model, e.g., through weight pruning, may improve the inference time.

Responsive design was rarely addressed in previous works. User interface designers typically deliver different designs for different screen sizes. Through the use of CSS media queries \cite[7 Media types]{bos_cascading_2016}, identical markup code can be delivered to various device types. A commercially valuable model should address this and accept multiple inputs for a single output. The published datasets (see section \ref{sec: Data}) contain vector images of multiple screen resolutions but during model training a single resolution was used, with the exception of a proof-of-concept experiment on synthetic data (see Appendix \ref{app:synthetic-data}).

Another problem often left unaddressed in image-to-code projects and benchmarks is the functional aspects of the generated code. For instance, without additional input, the model cannot infer whether a box in the design corresponds to a button or a text input. Accessibility and interactivity also remain unattended. The code quality of the resulting code is not a concern in current research.

\subsection{Future work}

More work is needed for high-accuracy, commercially viable image-to-code models. I believe that focusing on vector images over bitmaps is a promising research path.

Image-to-code problems employ rule-based verification (image similarity), which makes them a natural candidate for Reinforcement Learning with Verifiable Reward (RLVR) \cite{su_crossing_2025} algorithms such as GRPO \cite{shao_deepseekmath_2024}, GTPO \cite{tan_gtpo_2025}, or CAPO \cite{xie_capo_2025}.

More and higher-quality training data, as well as more modern base models, may further improve the accuracy. Training multimodal models on both vector and bitmap images of the same design is a potentially fruitful research direction as well.


\appendix

\section{Appendix}

\subsection{Synthetic data}
\label{app:synthetic-data}

\subsubsection{Datasets}

\begin{figure}[ht]
    \centering
    \centerline{\includegraphics[width=1.2\textwidth]{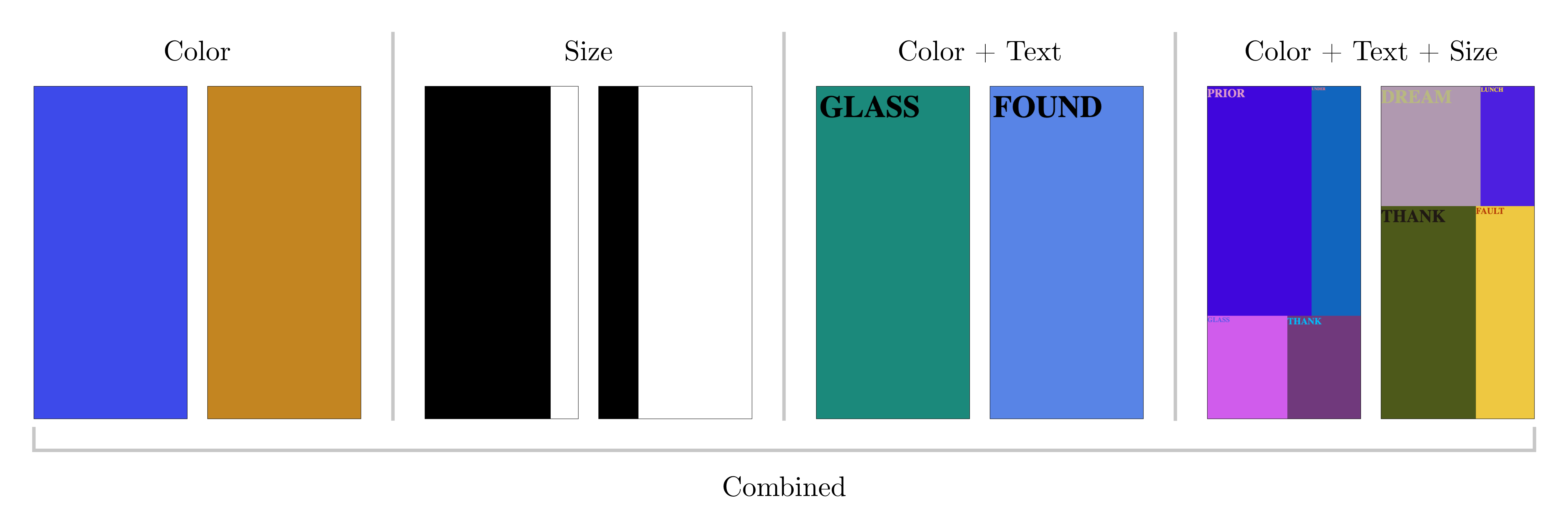}}
    \caption{Examples of web pages included in each simple synthetic dataset.}
    \label{fig:synthetic-data-set-screenshots}
\end{figure}

To prove that transformers can learn to translate between simple vector images and their corresponding web pages, I created a set of basic training datasets with increasing complexity. The web pages were generated using templates with placeholders for randomly chosen variables. \textcite{soselia_learning_2023} use a similar approach. Figure \ref{fig:synthetic-data-set-screenshots} shows example items of each dataset.

\paragraph{Color} The simplest possible web page with only a single CSS variable setting the page's background color.

\paragraph{Color Responsive} Like "Color" but the background differs on desktop and mobile screens.

\paragraph{Size} A dataset meant to test if the model can learn elementary arithmetic and translate between a percentage value and an absolute pixel value.

\paragraph{Color+Text} A dataset that extends the "Color" dataset by displaying a random word. 

\paragraph{Color+Text+Size} The most complex synthetic dataset where color, text, and size definitions are combined.

\paragraph{Combined} A heterogeneous dataset that combines all the above datasets (except "Color Responsive").

Synthetic datasets were published under the Apache 2.0 License at the following URLs.

\begin{table}[h]
    \begin{tabular}{ll}
    Color            & https://huggingface.co/datasets/tcz/rb-color \\
    Color responsive & https://huggingface.co/datasets/tcz/rb-color-responsive \\
    Size             & https://huggingface.co/datasets/tcz/rb-size \\
    Color+Text       & https://huggingface.co/datasets/tcz/rb-color-text \\
    Color+Text+Size  & https://huggingface.co/datasets/tcz/rb-color-text-size \\
    Small Validation & https://huggingface.co/datasets/tcz/rb-small-validation
    \end{tabular}
\end{table}

Table \ref{tab:synthetic-data-set-statistics} shows the sizes of the synthetic datasets.

\begin{table}[h]
\centering

    \begin{tabular}{lrrr}
     \hline
                     & Items & Total Tokens (millions) & Median Tokens Per Item \\ \hline
    Color            & 100,000   & 35 & 351 \\
    Color responsive & 100,512  & 71 & 709 \\
    Size             & 99,789 & 43 & 431 \\
    Color+Text       & 100,000 & 67 & 670 \\
    Color+Text+Size  & 99,853 & 217 & 2,178 \\  \hline                            
    Small Validation & 4 & $\approx 0$ & 1,406 \\  \hline       
    \end{tabular}

\caption{Total token sizes and median item token lengths of the synthetic datasets. Google's BERT multilingual base model tokenizer was used to calculate token counts. For the larger datasets, a sample of 5,000 random items was used to calculate token lengths.}
\label{tab:synthetic-data-set-statistics}
\end{table}

\subsubsection{Training}

To assess whether transformers are capable of learning to translate between vector images and markup, a series of training experiments was performed on the simple datasets described above. 

The T5 family (including LongT5 with an extended context window) was chosen as a cost-effective option where the context window was optionally easily expandable when needed. The T5 family is a sequence-to-sequence set of models, which seemed appropriate to implement an inherently sequence-to-sequence task, recognizing that the current state-of-the-art research is focused on autoregressive models and therefore the most powerful pre-trained models come from that class. A downside of T5 is that its vocabulary lacked some special markup characters, so it had to be extended before training. 

It was not clear whether this domain-specific task would benefit from a pre-trained model. The T5 family was trained on English text only, explicitly excluding source code in the training data. Therefore, nearly all the training experiments below were performed both on an untrained model with random initial weights (pre-training) and on the published pre-trained models (fine-tuning). Generally, the model converged much faster and stayed stable in later epochs when fine-tuned. This suggests transfer learning from a completely different domain.

\begin{table}[h]
    \centering
    \begin{tabular}{llrrrr}
        \hline
        Dataset name & Model name & \multicolumn{2}{c}{Pre-training} & \multicolumn{2}{c}{Fine-tuning} \\
        & & MSPS & LPIPS & MSPS & LPIPS \\ \hline
        "Color" & T5 small & 0.9642 & 0.0986 & 0.9935 & 0.0285 \\
        "Size" & T5 small & 0.8754 & 0.1526 & 0.9985 & 0.0318 \\
        "Color+Text" & LongT5 Base & 0.9725 & 0.0376 & 0.9982 & 0.0109 \\
        "Color+Text+Size" & LongT5 Base & 0.9998 & 0.0047 & 0.9999 & 0.0015 \\
        "Color Responsive" & LongT5 Base & 0.948  & 0.0989 & 0.9981 & 0.0152 \\
        "Combined" & LongT5 Base & N/A    & N/A    & 0.998  & 0.0212 \\
        \hline
    \end{tabular}
\end{table}

\subsection{On the Image2Struct benchmark}
\label{app:image2struct-more}

In this section, I explain additional details about how the Image2Struct modified benchmark seen in section \ref{sub: Benchmark} was performed. 

The Image2Struct \cite{roberts_image2struct_2024} web page dataset\footnote{https://huggingface.co/datasets/stanford-crfm/image2struct-webpage-v1} contains 900 web pages obtained from public GitHub Pages repositories. These are stored as individual ZIP files and include all the local assets (HTML, CSS, JavaScript, images) obtained from the repositories. The dataset is partitioned into three equally sized splits: HTML, CSS, and JavaScript. In my correspondence with the authors, they confirmed that the splits indicate that one of the three languages is more heavily represented in a given web page repository.

Image2Struct leaderboards show the combined score using the entire dataset and also publish separate scores for each split. I combined the splits into one dataset and only calculated the overall score. This also meant that the dataset had many JavaScript-heavy pages. Reverse Browser's renderer explicitly disables JavaScript in the crawled pages, and the model is not trained to generate JavaScript.

Image2Struct runs a local Jekyll\footnote{https://jekyllrb.com/} web server (also used by GitHub Pages) to serve the web pages and capture their content. I did the same, starting a separate server per page, and used the list of local URLs as input to my web page crawler described in Section \ref{sec: Data}.

Image2Struct captures the screen at $1920 \times 1080$ pixel resolution and discards all content outside the viewport when taking a screen capture. This is unlike the behavior of my crawler (which captures the entire page), so I made changes to follow this method. When crawling, I removed all DOM elements outside the viewport and captured SVG with only the remaining elements. I also saved all image resources that the web page loaded. These were made available later when rendering the pages generated by the model.

I then created a dataset with the obtained SVG-markup pairs. Here, I did not discard any items based on their MSPS score, resulting in a dataset of 884 items. Of the original 900 web pages, 16 had severe syntactic errors that prevented the SVG conversion.

I then performed inference on the final model using the SVGs and the same prompt I used during training. The notebook and the generated pages were published in the experiment repository\footnote{https://github.com/tcz/rb-experiments/blob/main/helm-generations/vllm-image2struct-inference.ipynb}. Image2struct uses a 0.0 temperature setting to avoid fluctuations in the performance. I followed the same strategy. They also do not use best-of-N prompting, nor did I. They have no documented mechanisms to catch the models "cheating", e.g., simply presenting a web page containing the image from the prompt. I also did not use any bad-word filters (as discussed in section \ref{sub: Use case}). Only a negligible percentage of the outputs contained text referring to SVG.

The final scores are based on 882 items as reflected in Table \ref{tab:image2struct-scores}. An additional two items were "lost" due to web page rendering problems with the output. The table separates \textit{Compilation success} as defined by Image2Struct and \textit{Training and inference success} that are influenced by different error modes as described here.

\setlength{\baselineskip}{0pt} 

\renewbibmacro*{urldate}{
(retrieved \printfield{urlday}/\printfield{urlmonth}/\printfield{urlyear})
} 

{\renewcommand*\MakeUppercase[1]{#1}%
\printbibliography[heading=bibintoc,title={References}]}

\end{document}